# All-magnonic repeater based on bistability

*Qi Wang[1], Roman Verba[2], Kristýna Davídková[3], Björn Heinz[4], Shixian Tian[5], Yiheng Rao[5], Mengying Guo[1], Xueyu Guo[1], Carsten Dubs[6], Philipp Pirro[4], Andrii V. Chumak[3]*

[1] *School of Physics, Huazhong University of Science and Technology, Wuhan, China*

[2] *Institute of Magnetism, Kyiv, Ukraine*

[3] *Faculty of Physics, University of Vienna, Vienna, Austria*

[4] *Fachbereich Physik and Landesforschungszentrum OPTIMAS, Rheinland-Pfälzische Technische Universität Kaiserlautern-Landau, Kaiserslautern, Germany*

[5] *School of Microelectronics, Hubei University, Wuhan, China*

[6] *INNOVENT e.V., Technologieentwicklung, Jena, Germany*

**Abstract:**

Bistability, a universal phenomenon found in diverse fields such as biology, chemistry, and physics, describes a scenario in which a system has two stable equilibrium states and resets to one of the two states. The ability to switch between these two states is the basis for a wide range of applications, particularly in memory and logic operations. Here, we present a universal approach to achieve bistable switching in magnonics, the field processing data using spin waves. As an exemplary application, we use magnonic bistability to experimentally demonstrate the still missing magnonic repeater. A pronounced bistable window is observed in a 1 μm wide magnonic conduit under an external rf drive characterized by two magnonic stable states defined as low and high spin-wave amplitudes. The switching between these two states is realized by another propagating spin wave sent into the rf driven region. This magnonic bistable switching is used to design the magnonic repeater, which receives the original decayed and distorted spin wave and regenerates a new spin wave with amplified amplitude and normalized phase. Our magnonic repeater is proposed to be installed at the inputs of each magnonic logic gate to overcome the spin-wave amplitude degradation and phase distortion during previous propagation and achieve integrated magnonic circuits or magnonic neuromorphic networks.

**Introduction**

A base station for mobile communications is one of a most common repeaters, i.e. a device that receives a signal, cleans it up, and then retransmits it at a required power level. Repeaters play a critical role in extending the range of signals, improving signal quality, and overcoming signal degradation that can occur over long distances, and are widely used in wireless, optical, and quantum communications [1-3]. Magnonics is an emerging field in which spin waves and their quantum magnons, the collective excitation of magnetic orders, are used for data transmission and processing. Recently, spin waves have attracted much attention in the field of conventional and unconventional computing [4-6] due to their nanoscale wavelengths [7-10], controllable nonlinear phenomena [11-14], energy efficiency [15] and the abundant interaction with other quasi-/particle [16-18]. Several individual magnon-based computing devices have been demonstrated including spin-wave logic gates [19], majority gates [20], magnon transistors [21-24], magnonic directional couplers [25], and neuromorphic computing elements [26]. However, an integrated magnonic circuit cascading multiple magnonic elements has not yet been realized experimentally due to the lack of the crucial repeater to overcome the degradation (decrease in amplitude) and distortion (deformed in phase) of the spin-wave signal during propagation.

Here, based on the recently discovered bistability of the deeply nonlinear forward volume spin waves excitation in nanoscale waveguides [14], we demonstrate an elegant way to switch between the two magnon states using a propagating spin wave, thus realizing a magnonic repeater. The repeater is a simple 2 μm wide strip antenna placed on top of a magnonic waveguide as shown in Fig. 1 and is proposed to be installed at the input of each magnonic logic gate to receive damped and distorted spin-wave signals from the upper-level logic gate, clean them up, and then regenerate new spin waves with amplified amplitude (with the gain up to 6 times) and normalized phase to connect the next level logic gate. This opens the door to cascading magnonic logic elements with amplitude information encoding, allowing the practical realization of complex Boolean circuits predicted in theory [27] as well as the realization of magnonic synapses in neuromorphic networks.

**Results**

**Concept and Working Principle of the Magnonic Repeater.** The main picture of Fig. 1 shows the schematic structure of the general concept of a magnonic repeater. A 1 μm wide yttrium iron garnet (YIG) waveguide is fabricated from a 44 nm thin film using a hard mask ion beam milling technique (see Methods) [14,28]. A coplanar waveguide (CPW) antenna and a 2 μm wide strip antenna are placed on top of the YIG waveguide with an edge-to-edge distance of about 5 μm. The bottom inset of Fig. 1 shows the scanning electron microscope (SEM) image of the experimental structure. An external field of 330 mT is applied out-of-plane along the z-axis and forward volume spin waves are investigated. Microfocused Brillouin light scattering spectroscopy (µBLS) is used to measure the spin-wave intensity at different positions along the YIG waveguide.

Figure 1 shows the schematic working principle of the magnonic repeater. Microwave pulses of the same frequency $f$ are sent to the two antennas. In principle, the excitation characteristics of both the antennas demonstrate bistable windows where the magnon state – low or high amplitude – depends on the prehistory, e.g. frequency sweep direction (Fig. 2). This bistability is a consequence of the large nonlinearity of forward spin waves in nanoscale waveguides [14] and is of similar nature as the bistability of a nonlinear oscillator when foldover effect takes place. At the same time, the bistable window for the used strip antenna is much wider due to the narrower k-spectrum of microwave field generated by this antenna.

This difference allows one to easily select the microwave frequency $f$ to ensure that the CPW antenna can directly excite the spin waves acting as a standard spin-wave source, and, at the same time, that the frequency $f$ is in the bistable window of the strip antenna. For the ground state (thermal level) of the magnonic waveguide, the microwave power of the strip antenna cannot be pumped into the magnonic domain due to the low excitation efficiency of the 2 µm wide antenna at the selected frequency $f$ and the large frequency gap $\Delta f$ between the excitation frequency $f$ and linear ferromagnetic resonance (FMR) frequency $f_{\text{FMR}}^{\text{lin}}$. However, when the propagating spin waves reach the strip antenna, they create a new "starting state" with nonvanishing amplitude, and if this amplitude is large enough (see criterion below), the magnon state under the strip antenna develops into the high-amplitude state. The input spin wave thus acts as a trigger allowing the microwave energy in the strip antenna to be pumped into the magnonic domain, exciting new spin waves and completing the magnonic bistable switching. More interestingly, the phase of the newly excited spin waves is determined by the phase of the microwave signal at the strip antenna, while their amplitude is purely determined by the selected working frequency due to the self-normalized excitation [14]. Thus, the original signal has been cleaned up by the re-emitting process, which improves the quality of the retransmitted signal and allows also for amplification.

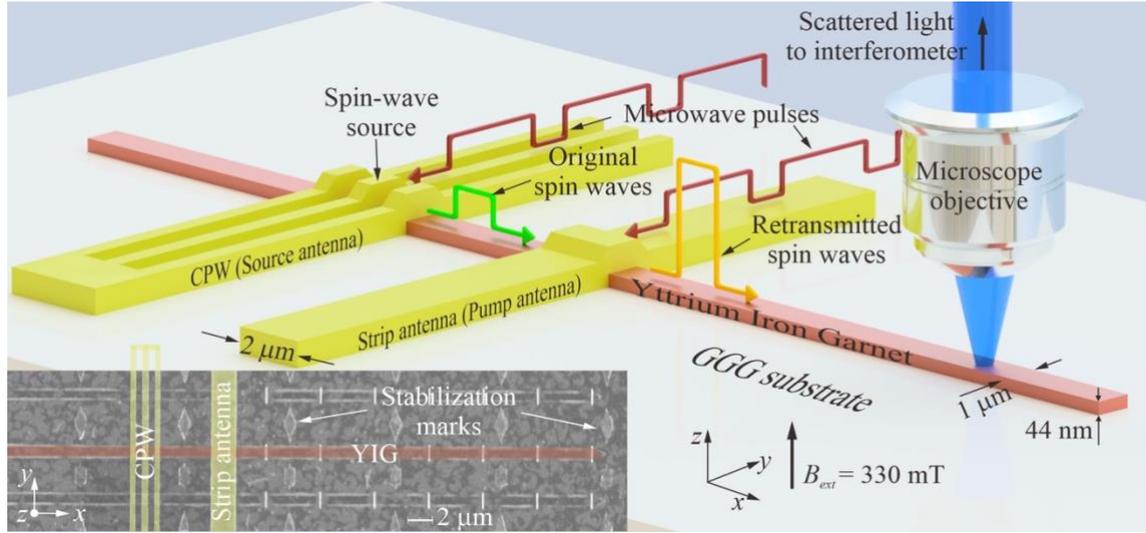

*Figure 1. **The structure of a magnonic repeater.** Sketch of the sample and the experimental configuration: a CPW antenna (source antenna) is placed on a 1 μm wide yttrium iron garnet (YIG) waveguide to directly excite spin waves, which acts as a spin-wave source. A 2 μm wide strip antenna (pump antenna) is used to receive original spin waves and regenerate new spin waves working as a repeater. μBLS spectroscopy is employed to detect the spin-wave intensity in the YIG waveguide at different positions. The left bottom inset shows the SEM image of the experimental structure. The yellow areas indicate the CPW and strip antennas. The dark red part shows the YIG waveguide. The rhombic markers are used to stabilize the sample during the measurements. The vertical nanowires on the surface of the magnonic waveguide are used to improve the detection efficiency of short wavelengths and are not used in this work. GGG, gadolinium gallium garnet.*

**Bistable window.** The curve of figure 2 was obtained by sending a continuous microwave current with a power of 6 dBm, swept from 4.7 GHz to 8 GHz (or vice versa) with a step size of 20 MHz, to the 2 μm wide strip antenna to excite spin waves, and the focused laser spot of the μBLS is placed about 4 μm from the edge of the strip antenna to detect the excited propagating spin waves. The frequency step is small enough so that magnon state evolves from a preceding one. As a result, the up and down frequency sweep curves do not overlap and show a hysteretic response resulting in a large bistable frequency window of ~1.1 GHz which is similar to our previous study [14]. The nature of this bistable window is similar to those observed in a common nonlinear oscillator (e.g., Duffing oscillator) when the foldover effect takes place, although there are certain differences (see Supplementary Materials). In addition, Fig. 2 shows several small peaks around 6 GHz, 6.6 GHz and 7.2 GHz, which belong to the higher-order spin-wave width modes that are excited by the 2 μm wide antenna with two orders of magnitude lower amplitude and, therefore, can be ignored in our studies. The simulated snapshot of the propagating spin waves and the BLS measured profile across the waveguide clearly show that only the fundamental mode is excited in the 1 μm wide waveguide at the highest state within the bistable window (see the Supplementary Materials).

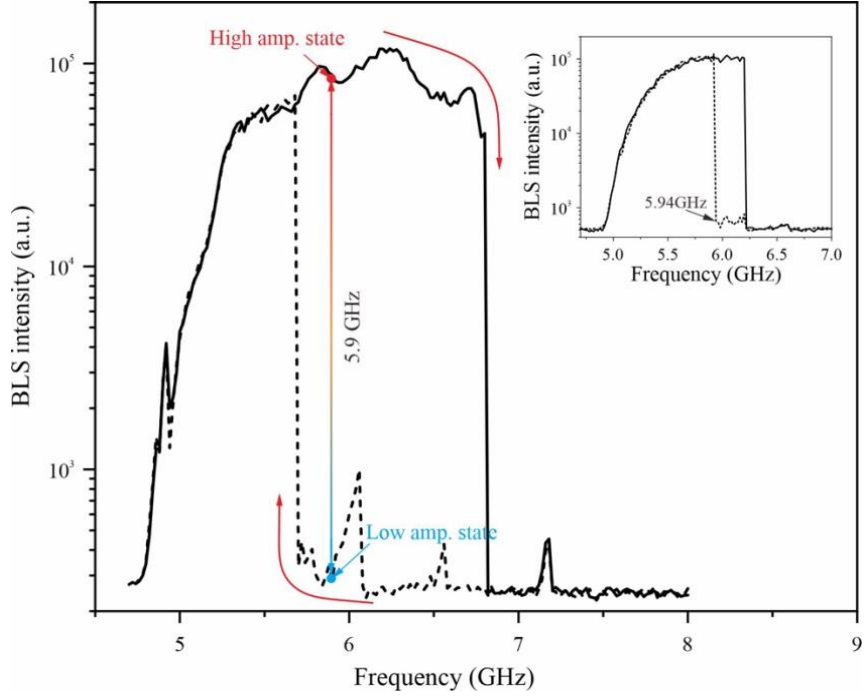

*Figure 2. **Magnonic bistable window with two equilibrium states.** μBLS intensity as a function of excitation frequency f at a microwave power of P=6 dBm applied to a 2 μm wide strip antenna for up (solid line) and down (dashed line) frequency sweep, respectively. The frequency f=5.9 GHz is used to demonstrate magnonic bistable switching. It is located in the bistable window and the corresponding two stable states are marked as high/low amplitude state. The inset shows a similar spectrum, but excited by the CPW antenna. For this antenna, f=5.9 GHz is located at outside of the left edge of the bistability window and can be excited without hysteretic effects.*

**Bistable Switching and Magnonic Repeater.** To utilize the bistability for designing a magnonic repeater, one needs an efficient and unambiguous method for the switching between the two states in the bistable window. Sure, an adiabatic frequency sweep, used for measurements of the excitation spectra Fig. 2, cannot be a method of choice. That is why we use the CPW antenna introduced in Fig. 1 to excite propagating spin waves as a source to switch the bistable states under the stripe antenna.

Figure 3 shows the general working principle of magnonic bistable switching for the realization of a magnonic repeater. A frequency of 5.9 GHz within the bistable window as shown in Fig. 2 was exemplary chosen to demonstrate the functionality of the switching. First, we apply microwave pulses of this frequency with a duration of 500 ns and a repetition time of 1 μs at a power of 6 dBm to the 2 μm strip antenna (pump antenna) only, as shown in the first column of Fig. 3(a). Time-resolved microfocused Brillouin light scattering spectroscopy (μBLS) is used to measure the spin-wave intensity as function of time in the frequency range from 3 GHz to 9 GHz. The laser spot is placed about 4 μm from the edge of the pump antenna as marked by blue dot. The second column of Fig. 3(a) shows the two-dimensional color-coded magnon spectra as a function of time, where the BLS signal (color-coded, log scale) is proportional to spin-wave intensity. Two faint horizontal lines around 4 GHz and 8 GHz are the laser side modes. No other signals indicating the excitation of spin waves are observed. A similar result is shown in the third column of Fig. 3(a), where the integrated signal in the range 5-7 GHz (between the two horizontal dashed lines) is plotted. In this case, the initial state is the low-amplitude thermal background, and the system falls into the low-amplitude state, where the spin-wave amplitude is very close to the thermal level (see Fig. 2) and cannot be detected by time-resolved BLS.

Figure 3(b) shows the case where the same microwave pulses are sent only to the CPW antenna (source antenna). Propagating spin-wave signals are observed in the time-dependent magnon spectrum and the integrated signal, in full accordance with the inset of Fig. 2 - due to the high excitation efficiency at nonzero wavenumbers of the CPW antenna, it excites spin waves at 5.9 GHz independently on the prehistory, i.e. without hysteresis and bistability (see the inset of Fig. 2) [29].

Figure 3(c) shows the case when microwave pulses were sent simultaneously to the source and pump antennas. It can be clearly seen from the time evolution of the magnon spectrum that the magnon intensity is much stronger and the thermal background has a similar noise level compared to the case of the CPW source alone, indicating that only the propagating spin waves were repeated by the pump antenna. The small tails observed at the end of each pulse are caused by the nonlinear self-phase modulation [30].

Figure 3(d) shows the case where the source and pump pulses have different periods to ensure that every second spin-wave pulse is reemitted by the pump antenna. It is clear from the integrated signal that the amplification in this case is about six times. The influences of the source and the pump powers are discussed in the supplementary materials, showing that the proposed method is robust with a large working window. Furthermore, the experimental results are verified by micromagnetic simulations.

The physics behind the operation of the repeater is the nonlinear frequency shift of spin waves, which is described as

$$f_k(c_k) = f_{k,0} + T_k c_k^2 \qquad (1)$$

where $f_{k,0}$ is the linear (small-amplitude) spin-wave frequency, $T_k > 0$ is the nonlinear frequency shift coefficient, and $c_k$ is canonic spin wave amplitude. In our case of forward spin waves, $c_k^2 \approx 1 - \overline{\cos\theta}$, where $\theta$ is the precession angle and overbar means averaging over the waveguide width [14]. The nonlinear frequency shift coefficient in the range from $k = 0$ to $k \approx 20\,\mu m^{-1}$, which is the wavenumber of spin waves excited by the CPW antenna, weakly depends on $k$. Then, the effect of the incident spin waves can be imagined as a shift of the whole spin wave spectrum by the value $\Delta f = T_k c_k^2$ [31]. The excitation characteristic of the pump antenna (Fig. 2) is, thus, also shifted by the same value $\Delta f$. If this shift is greater than the distance between the working frequency and low-frequency edge of the bistability window, the working frequency $f$ becomes resonant in this new characteristic, shifted by the presence of incident spin waves. Thus, the pump antenna excites spin waves at the frequency $f$, which can be of sufficiently larger magnitude than incident spin wave.

Numerical values, obtained from analytic calculations and micromagnetic simulations, are in full accordance with the above picture. In the presented case, the low-frequency edge of the bistability window is located at approximately 5.7 GHz, thus, 200 MHz below the working frequency of 5.9 GHz. The nonlinear frequency shift coefficient for $k \approx 20\,\mu m^{-1}$ is $T_k \approx 6.35$ GHz (see calculation method in [14]). In the simulations, a little above the switching threshold, the propagating spin waves arriving at the pump antenna have the width-averaged precession angle of $\bar\theta = 12.3°$, which corresponds to canonic spin wave amplitude $c_k \approx 0.185$. Thus, incoming spin waves shift the excitation characteristic by $\Delta f \approx 220$ MHz, which is larger than the 200 MHz gap between the working frequency and thus enough to switch the pump antenna into the excitation state with a large-amplitude magnon.

This consideration shows that the threshold power of the incident spin waves required for switching the pump antenna into the high amplitude state is proportional to the distance between the operating frequency and the left edge of the bistability window $f_{bs}$, $c_{k,th}^2 \sim (f - f_{bs})$. It is also expected that the power and the phase of the repeated pulse is completely determined by the microwave pulse submitted to the pump antenna. Indeed, the role of the incident spin wave is just switching of the pump antenna in the bistable window from the vanishing-amplitude state to the large-amplitude state. As soon as it has been switched, the pump antenna remains in the high-amplitude state even if the incoming spin wave pulse falls below the threshold amplitude.

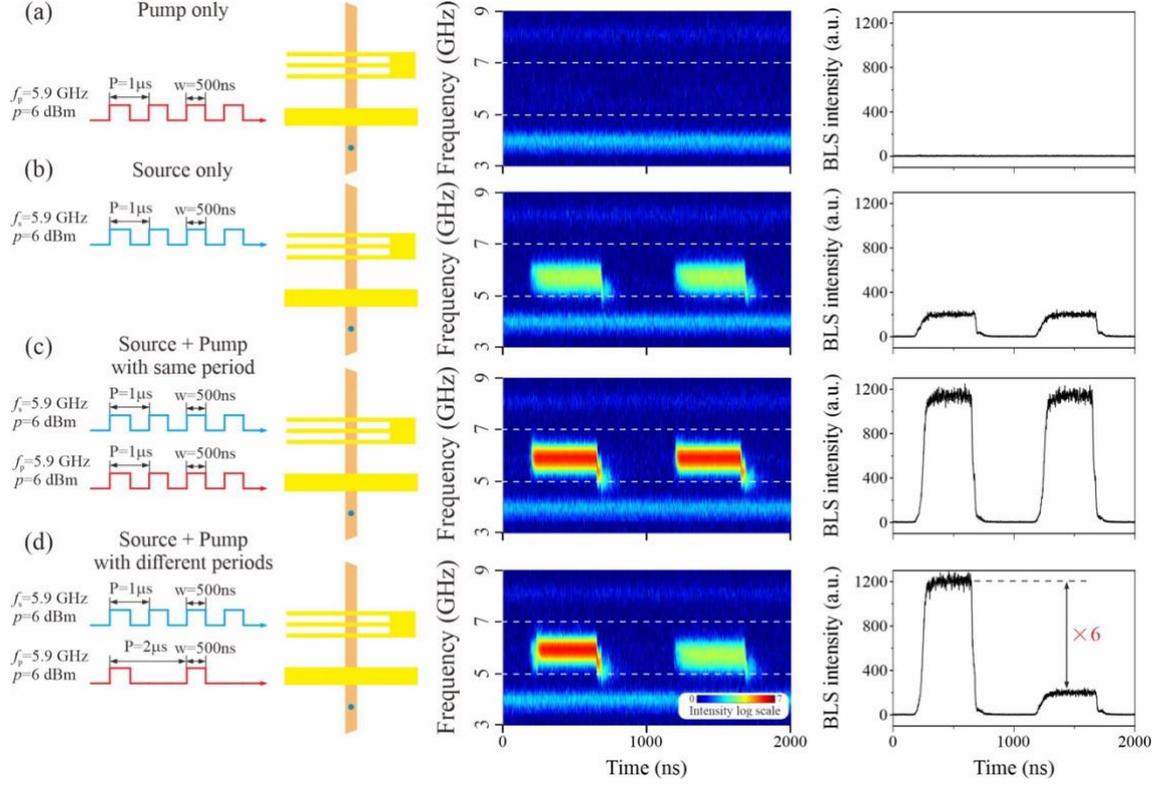

*Figure 3. **Working principle of the magnonic repeater.** (a) Pump only, (b) Source only, (c) Source + pump with same period, (d) Source + pump with different periods. The first column shows the schematic pictures for four different cases. The blue dots in the sketch indicate the (fixed) laser spot position of the BLS during the experimental measurements. The second column shows the two-dimensional BLS spectra as a function of time. The BLS signal (color-coded, log scale) is proportional to the intensity of the magnons. All spectra share the same color code. The third column shows the BLS intensity as a function of time, integrated from 5 GHz to 7 GHz, as indicated by the two horizontal dashed lines in the two-dimensional BLS spectra.*

**Amplitude and Phase Normalization.** In the following, we examine the amplitude and phase characteristics of the retransmitted spin waves. Figure 4(a) illustrates the output spin-wave intensity as a function of the source power at a pump power of 9 dBm. It is noticeable that once the source power exceeds a certain threshold (approximately 6 dBm in our case), the output spin-wave intensity is nearly constant and is also insensitive to the pump power, as we expected from the above physical picture (see Supplementary Materials). In terms of phase information, as discussed above, the amplification is attributed to the regenerated higher amplitude spin waves by the pump antenna. Thus, the phase of the regenerated spin waves is aligned with the phase of the microwave in the pump antenna, as demonstrated in Fig. 4(b). This amplitude and phase normalization offers remarkable advantages, making the device into an efficient repeater to overcome the spin-wave amplitude degradation and phase distortion during its propagation. Therefore, the subsequent logic gates can directly utilize the retransmitted spin waves from the magnonic repeater as input signals without the need for further phase or amplitude modulation.

Finally, in order to prove its universality, additional micromagnetic simulations were performed with different materials (e.g., CoFeB) and varying sizes (down to 100 nm waveguide width) (see the Supplementary Materials). The consistent reproduction of the repeater behavior across different materials and sizes confirms its potential as a universal method.

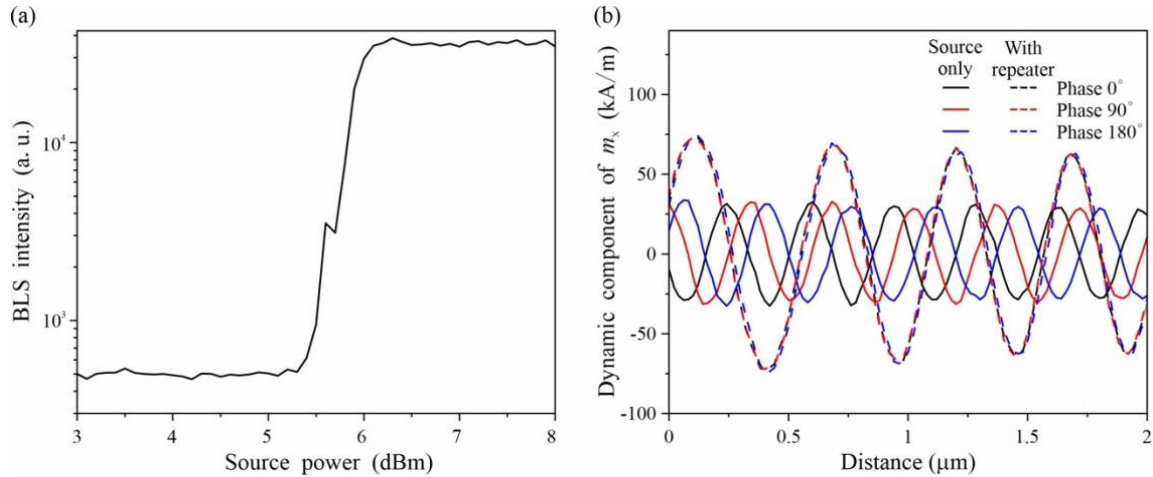

*Figure 4. **Amplitude and phase normalization**. (a) The experimental results of the output spin-wave intensity as a function of the spin-wave source power. (b) The simulated spin-wave amplitude at different source phases with repeater (dashed lines) and source only (solid lines).*

**Discussion**

We have experimentally observed a large bistable window of 1.1 GHz in a 1 μm-wide YIG waveguide subjected to a rf magnetic field via a strip antenna. This pump antenna provides two stable magnon states with high and low spin-wave amplitudes and allows us to switch between them. In addition, we have placed a second antenna on top of the magnonic waveguide to act as source. Spin-wave pulses emitted by the source antenna can be reemitted by the pump antenna. This is possible since the incident propagating spin waves shift the excitation characteristic of the pump antenna via the nonlinear frequency shift of spin wave spectrum. This enables to switch from the low-amplitude state in the bistable window to the high-amplitude state. This switching results in an amplification of the retransmitted spin wave as compared to the incident one, which reaches 6 times in our experiments. In addition, the amplitude and phase of the output signal are independent of the source and pump power allowing for a simplified and robust design of magnonic circuits. Further simulations reveal that this mechanism is a universal and robust method, suitable for other materials, and can be scaled down to tens of nanometers for nanoscale circuits. Looking to the future, we would like to emphasize that the presented magnonic bistability switching can also be used for other promising applications such as neural networks and stochastic computation.

## Methods

**Nanoscale waveguide and antenna fabrication.** The YIG thin film is grown on top of a 500 μm thick (111) gadolinium gallium garnet (GGG) substrate by liquid phase epitaxy (LPE) [32]. The parameters of the unstructured thin film were characterized by stripline vector network analyzer ferromagnetic resonance spectroscopy and BLS spectroscopy and obtain a saturation magnetization of $M_s = (140.7 \pm 2.8)$ kA/m, Gilbert damping parameter $\alpha = (1.75 \pm 0.08) \times 10^{-4}$, inhomogeneous linewidth broadening $\mu_0 \Delta H_0 = (0.18 \pm 0.01)$ mT, and exchange constant $A_{ex} = (4.22 \pm 0.21)$ pJ/m. These parameters are typical for high quality thin YIG films [28,32]. Nanoscale YIG waveguides were fabricated using a Cr/Ti hard mask and ion beam milling process, as described in detail in Ref. [14,28]. The CPW antenna of a ground-signal-ground line width of 400 nm-600 nm-400 nm and an edge-to-edge spacing of 600 nm is fabricated together with a 2 μm antenna using a typical electron beam lithography technology.

**BLS measurements.** A single-frequency laser with a wavelength of 457 nm is used, focused on the sample using a microscope objective (magnification 100× and numerical aperture N.A.=0.75). The laser power of 2.8 mW is focused on the sample. A uniform out-of-plane external field of 330 mT is provided by a NdFeB permanent magnet with a diameter of 70 mm. Microwave signals with different powers and frequencies were applied to the antenna to excite and retransmit spin waves.

**Micromagnetic simulations.** The micromagnetic simulations were performed by the GPU-accelerated simulation package Mumax$^3$, including both exchange and dipolar interactions, to calculate the space- and time-dependent magnetisation dynamics in the investigated structures [33]. The parameters of a nanometre-thick YIG film were used [14,28]: saturation magnetisation $M_s = 1.407 \times 10^5$ A/m, exchange constant $A = 4.2$ pJ/m. The Gilbert damping is increased to $\alpha = 5 \times 10^{-4}$ to account for the inhomogeneous linewidth which cannot be directly plugged into Mumax$^3$ simulations. The Gilbert damping at the end of the device was set to exponentially increase to 0.5 to avoid spin-wave reflection. The mesh was set to 20×20×44 nm$^3$ (single layer along the thickness) for the YIG waveguide. An external field $B_{ext} = 330$ mT is applied along the out-of-plane axis (z-axis as shown in Fig. 1) and thus sufficient to saturate the structure in this direction.

The typical parameters of CoFeB were used: saturation magnetisation $M_s = 12.5 \times 10^5$ A/m, exchange constant $A = 15$ pJ/m, and the Gilbert damping $\alpha = 2 \times 10^{-3}$. The mesh was set to 5×5×5 nm$^3$. An external field $B_{ext} = 2.55$ T is applied to out-of-plane.

To excite propagating spin waves, we first calculate the Oersted field distribution of a 2 μm wide strip antenna with current of 25 mA in the magneto-static approximation and plug it into Mumax$^3$ with a varying microwave frequency $f$. The $M_y(x,y,t)$ of each cell was collected over a period of 100 ns and recorded in 25 ps intervals. The fluctuations $m_y(x,y,t)$ were calculated for all cells via $m_y(x,y,t) = M_y(x,y,t) - M_y(x,y,0)$, where $M_y(x,y,0)$ corresponds to the ground state. The spin-wave dispersion curves were calculated by performing a two-dimensional fast Fourier transformation of the fluctuations.

**Acknowledgements**

**Funding:** The project is funded by the National Key Research and Development Program of China (Grant No. 2023YFA1406600), the Austrian Science Fund (FWF) via Grant No. I 4696-N (Nano-YIG) and Grant No. F65 (SFB PDE), the European Research Council (ERC) Starting Grant 678309 MagnonCircuits and ERC Starting Grant 101042439 "CoSpiN" and the Deutsche Forschungsgemeinschaft (DFG, German Research Foundation) – 271741898 and TRR 173 - 268565370 ("Spin + X", Project B01). Q. W. was supported by the startup grant of Huazhong University of Science and Technology Grants No. 3034012104. R.V. acknowledges support by the Ministry of Education and Science of Ukraine, project # 0124U000270 and by IEEE via "Magnetism in Ukraine Initiative" (STCU project No. 9918). We thank Ondřej Wojewoda and Michal Urbánek for the fruitful discussion on the BLS measurements.

**Author contributions:** Q. W. proposed the magnonic repeater design and performed BLS measurements with help from M. G. and X. G. A. V. C. and P. P. led this project. R. V. provided theoretical support and analysis. B. H. fabricated the nanoscale YIG waveguides with help from K. D. Q. W. performed the micromagnetic simulations with help from S. T. and Y. R. C. D. grew the YIG film. Q. W. wrote the manuscript with the help of all the coauthors. All authors contributed to the scientific discussion and commented on the manuscript.

**Competing interests:** The authors declare no competing interests.

**Data availability:** The data that support the plots presented in this paper are available from the corresponding authors upon reasonable request.

**Code availability**

The code used to analyze the data and the related simulation files are available from the corresponding author upon reasonable request.

# Supplementary Materials
# All-magnonic repeater based on bistability

Qi Wang[1], Roman Verba[2], Kristýna Davídková[3], Björn Heinz[4], Shixian Tian[5], Yiheng Rao[5], Mengying Guo[1], Xueyu Guo[1], Carsten Dubs[6], Philipp Pirro[4], Andrii V. Chumak[3]

[1] School of Physics, Huazhong University of Science and Technology, Wuhan, China

[2] Institute of Magnetism, Kyiv, Ukraine

[3] Faculty of Physics, University of Vienna, Vienna, Austria

[4] Fachbereich Physik and Landesforschungszentrum OPTIMAS, Rheinland-Pfälzische Technische Universität Kaiserlautern-Landau, Kaiserslautern, Germany

[5] School of Microelectronics, Hubei University, Wuhan, China

[6] INNOVENT e.V., Technologieentwicklung, Jena, Germany

1. **Simulated Foldover and Bistability**

In the main manuscript, there are several small peaks in the BLS spectra visible (see Fig. 2), which are attributed to the higher width modes and can be ignored in our studies. To verify this, we performed micromagnetic simulations similar to the experiments, where the excitation frequency was swept from 4.3 GHz to 8.0 GHz (or vice versa) with a step size of 50 MHz. To excite spin waves, we first calculate the Oersted field distribution of a 2 µm wide strip antenna with current of 20 mA in the magneto-static approximation and plug it into Mumax$^3$ with a varying microwave frequency $f$. The $M_y(x,y,t)$ of each cell was collected over a period of 100 ns and recorded in 50 ps intervals. The oscillations $m_y(x,y,t)$ were calculated for all cells via $m_y(x,y,t) = M_y(x,y,t) - M_y(x,y,0)$, where $M_y(x,y,0)$ corresponds to the ground state. The spin-wave intensity was extracted 5 µm far from the antenna. All the simulations were performed with a temperature of 300 K.

Figure S1(a) shows the simulated spin-wave spectra with a foldover effect and bistable window similar to the experimental results. Several small peaks are also observed in the simulations corresponding to the higher width modes with one or two orders of magnitudes smaller intensity. Interestingly, the observed higher-order modes are the 7$^{th}$, 9$^{th}$ and 11$^{th}$ modes, as confirmed by the comparison with a linear excitation spectrum at 10 times smaller driving field (see the red dashed line in Fig. S1(a)). Simultaneously, in the nonlinear excitation regime we don't observe the excitation of 3$^{rd}$ and 5$^{th}$ modes (even modes have zero overlap with rf field and cannot be excited).

We verified the excitation of the fundamental mode in the nonlinear regime. Figure S1(b) show the simulated spin-wave amplitude distribution for a frequency of 6.9 GHz along the waveguide for the case of the frequency upsweep. It is clear that only the fundamental width mode is excited. A corresponding experiment is performed by sending a microwave current of frequency 6.9 GHz at power of 15 dBm to the pump antenna to directly excite propagation spin waves. The BLS is used to measure the spin-wave intensity across the width of the waveguide as shown in Fig. S1(c). A sinusoidal shape of the profile is obtained indicating the fundamental mode. The same, fundamental mode was the only observed at other points at the frequency upsweep path until 7.5 GHz, where the excitation drops.

Suppression of higher-order modes in the nonlinear regime is quite simple in nature. As explained in [S1], under a deeply nonlinear excitation, the nonlinear ferromagnetic resonance frequency under the antenna is shifted to match the excitation frequency. Similarly, the frequencies of the higher-order width modes also shift up, and the whole higher-order branch lies above the excitation frequency. Thus, only the fundamental mode survives. This is another intriguing advantage of the deeply nonlinear excitation mechanism. In contrast, in the linear mode, multiple mode can be excited in a wide waveguide.

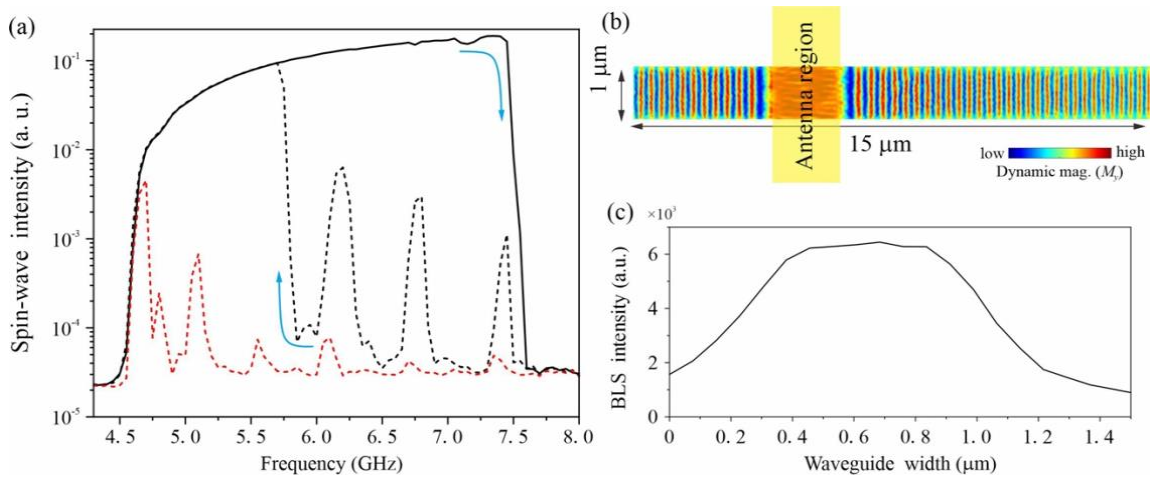

*Fig. S1 **Simulated Foldover effect and Bistability**. (a) Simulated spectra show a foldover effect and large bistable window. Red dashed line shows the down-sweep curve with 10 times smaller driving field. (b) The simulated spin-wave amplitude distribution (frequency of 6.9 GHz) along the waveguide for the case of up frequency sweep. (c) The BLS intensity of frequency 6.9 GHz (Power P=15 dBm) across the width of 1 μm wide waveguide.*

**2. Influence of the Source and Pump Power**

Figure S2 shows the influence of source and pump power on the spin-wave intensity for the case of source + pump with the same period as shown in Fig. 3(c) in the main manuscript. Figure S2(a) and (b) show the experimental results of the integrated spin-wave intensity as function of time for different (a) source powers (pump power @10 dBm) and (b) pump powers (source power @6 dBm). It can be seen that the spin-wave intensity is almost constant once the source or pump power are above their respective power thresholds, indicating a stable and large operating window. Only the switching time changes as the source and pump powers are varied. The simulated results shown in Fig. S2(c) and (d) reproduce this phenomenon well.

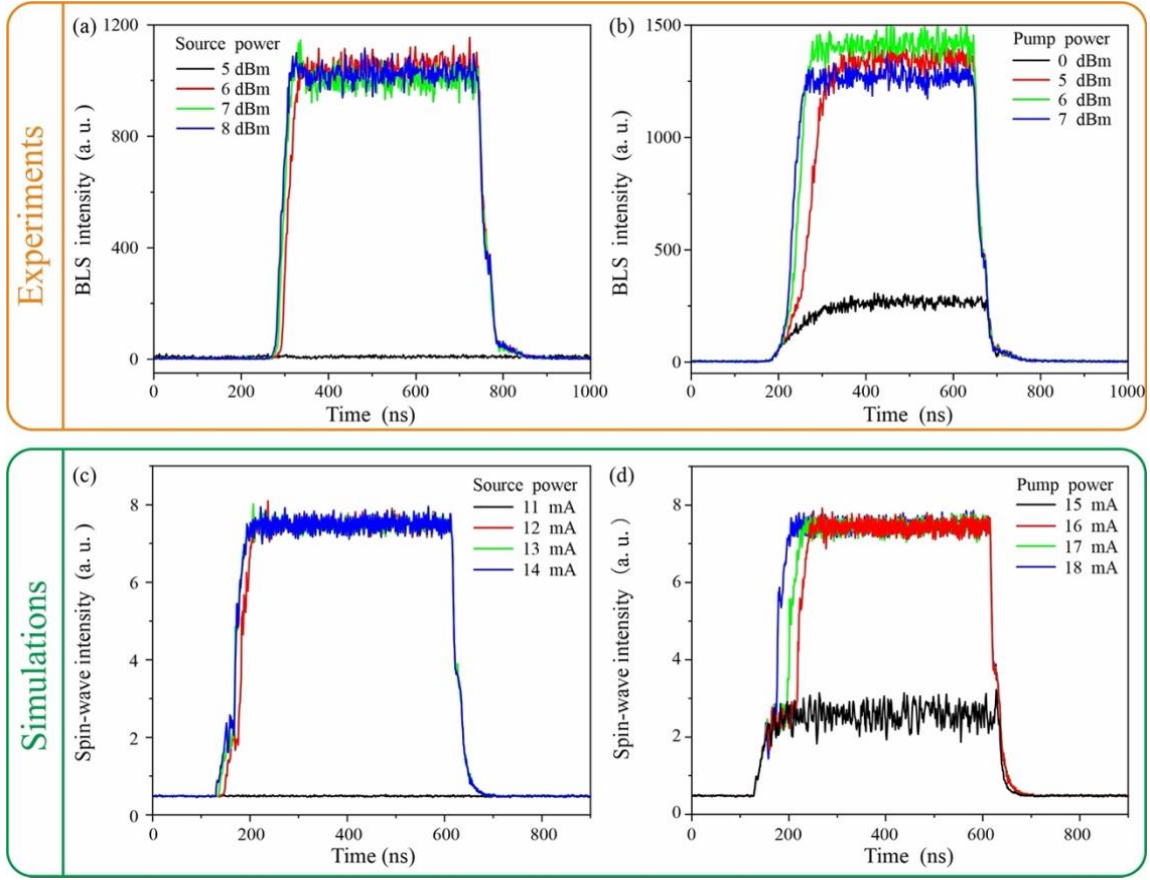

*Figure S2. **Influence of source and pump power.** The integrated spin-wave intensity as function of time for different (a) source powers (pump power @10 dBm) and (b) pump powers (source power @6 dBm). The simulated spin-wave intensity for different (c) source powers (pump power @20 mA) and (d) pump powers (source power @15 mA).*

### 3.  Magnonic Repeater with Different Materials and Sizes

In order to prove its universality, additional micromagnetic simulations were performed with different materials and varying size. Figure S3 shows the simulated spin-wave intensity of three different cases: source only, pump only and source + pump for (a) 100 nm wide and 44 nm thick YIG waveguide with working frequency of 8.8 GHz, (b) 50 nm wide and 5 nm thick CoFeB waveguide with working frequency of 61 GHz. The results are similar to Fig. 3 in the main text: the spin-wave intensity is amplified once the source and pump are simultaneously applied which indicates that the bistable switching using a propagation spin wave is a universal method in magnonics for different materials and varying sizes.

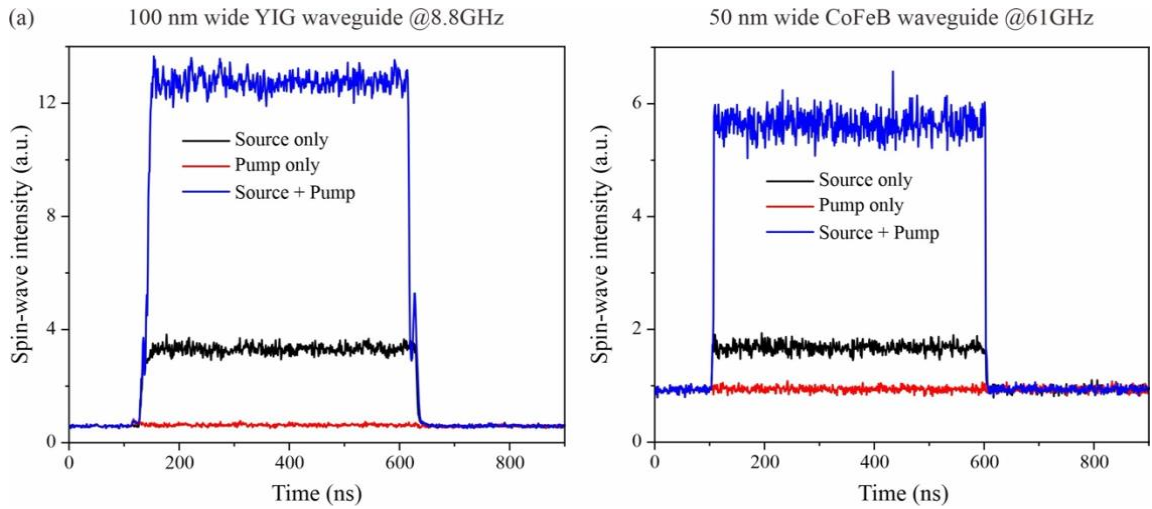

*Fig. S3 **Influence of different materials and varying sizes.** The simulated spin-wave intensity of three different cases: source only, pump only and source + pump for (a) 100 nm wide and 44 nm thick YIG waveguide with excitation frequency of 8.8 GHz, (b) 50 nm wide and 5 nm thick CoFeB waveguide with excitation frequency of 61 GHz.*